\documentclass{vgtc}      
\usepackage{multirow}

\graphicspath{{figures/}{pictures/}{images/}{./}} 

\usepackage{times}                     

\usepackage{mathptmx}                  
\usepackage{amsmath}
\usepackage{enumitem}
\setlist{topsep=0pt, leftmargin=*}

\onlineid{2064}

\vgtccategory{Research}

\vgtcinsertpkg

\title{Safe Guard: an LLM-agent for Real-time Voice-based Hate Speech Detection in Social Virtual Reality}

\author{Yiwen Xu\thanks{e-mail: xu.yiwen1@northeastern.edu}\\ %
\and Qinyang Hou\thanks{e-mail: hou.q@northeastern.edu}\\ %
\and Hongyu Wan\thanks{e-mail: wan.hongy@northeastern.edu}\\ %
\and Mirjana Prpa\thanks{m.prpa@northeastern.edu}}%
\affiliation{\scriptsize Northeastern University, Khoury
College of CS
Vancouver, Canada}

\teaser{
  \centering
  \includegraphics[width=\linewidth]{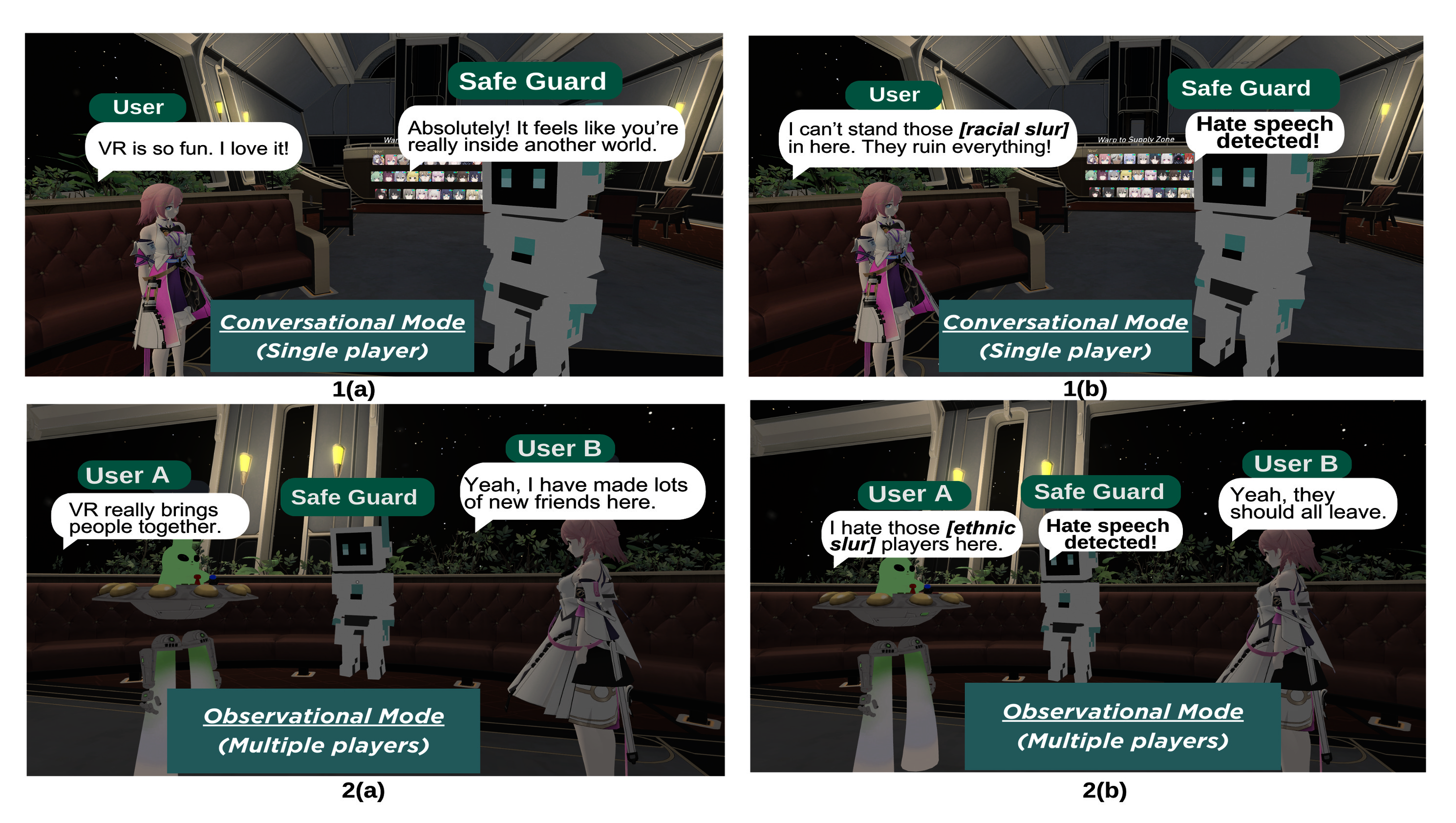}
  \caption{1(a) shows an interaction between the LLM-based agent \textit{Safe Guard} and a single player in VRChat in conversational mode. 1(b) shows an instance where hate speech is detected in conversational mode. 2(a) shows an interaction between LLM-based agent \textit{Safe Guard} and multiple players in VRChat in observational mode. 2(b) shows hate speech being detected in observational mode.}
  \label{fig: teaser}
}

\abstract{
In this paper, we present \textit{Safe Guard}, an LLM-agent for the detection of hate speech in voice-based interactions in social VR (VRChat). Our system leverages Open AI GPT and audio feature extraction for real-time voice interactions. We contribute a system design and evaluation of the system that demonstrates the capability of our approach in detecting hate speech, and reducing false positives compared to currently available approaches. Our results indicate the potential of LLM-based agents in creating safer virtual environments and set the groundwork for further advancements in LLM-driven moderation approaches.
} 


\begin{document}


\firstsection{Introduction}

\maketitle

Social Virtual Reality (VR) platforms such as VRChat, Rec Room, Bigscreen, AltspaceVR, and Meta Horizon Worlds \cite{Schulenberg:2023:AI_Moderation} have gone through a substantial rise in popularity in recent years. These platforms provide integrated features such as customizable avatars and real-time voice chat, offering users immersive first-person online social experience where they can communicate, interact, and engage in a more immersive and embodied way using head-mounted displays (HMDs) compared to traditional online platforms \cite{Zheng:2023:SRS}. In social VR, real-time voice interactions are fundamental. Users engage in conversations using their real voices and communicate with others in real time (RT) within the virtual environment. This form of interaction enhances users' feeling of presence and connection, making the experience resemble real-life face-to-face communication compared to text-based interactions \cite{Schulenberg:2023:AI_Moderation}.

However, as the increasing number of users engage in voice interactions within these virtual environments, new challenges and risks continue to arise in ensuring safe and trustworthy communication in social VR settings. Hate speech, as one of the main harassment forms on social VR platforms, poses significant threats to user well-being and the overall health of the VR communities. People who encounter online harassment often report significant disruptions to their offline lives \cite{Blackwell:2017:COH}, encompassing emotional and physical distress, shifts in their future technology usage, and raised concerns about safety and privacy \cite{Freeman:2022:DTP}. Zheng et al. \cite{Zheng:2023:SRS} discussed that online harassment within immersive environments like social VR settings could potentially cause more serious and negative effects on the target individuals’ well-being than traditional online attacks. Consequently, effective hate speech detection mechanisms are indispensable for social VR to mitigate the risks and maintain positive user experiences. 

Combating hate speech in social VR faces several challenges. Social VR safety risks are often unpredictable, highly personal, and real-time, making documenting and data collecting difficult \cite{Zheng:2023:SRS}. The immersive nature of social VR and the immediacy of voice interactions make it difficult for the platforms to detect and moderate toxic behaviors effectively \cite{Zheng:2023:SRS}. To this date, the most accurate method for combating hate speech is human moderation conducted by the community members, yet with the rise of the number of events and attendees, moderators are presented with scalability challenges due to the limited number of moderators and the burden that moderation poses on community members \cite{Schulenberg:2023:AI_Moderation}.

To overcome the challenges, we draw on the work on AI-moderation of harassment in social VR by Schulenberg et al. \cite{Schulenberg:2023:AI_Moderation} and a study conducted by Fiani et al. \cite{Fiani:2023:BB} which proposed the design direction for embodied AI moderators in social VR to mitigate harassment towards children. These studies highlighted the potential of AI agents in improving the moderation process by providing timely and context-aware interventions, which is crucial in dynamic and immersive settings like social VR. Building on the insights, our study focuses on the development of an LLM agent named "\textit{Safe Guard}" to detect real-time voice-based hate speech in social VR environments (see Figure \ref{fig: teaser}). Recent advancements in Large Language Models (LLMs) demonstrate under-explored potential for the use of LLMs in the context of real-time voice-based moderation in VR. To that end, we leveraged GPT-3.5 into our LLM-agent design for hate speech detection in VRChat. The proposed agent has two modes: (1) conversational mode with a single user, where it conducts conversation while simultaneously detecting and alerting for hate speech (2) observational mode when multiple users are present, where it monitors interactions to detect and alert for hate speech. 

LLMs present promising solutions for enhancing moderation capabilities in VR settings, by exhibiting the capabilities to effectively identify and classify hate speech based on context and textual content. Kolla et al. \cite{Kolla:2024:LLM} highlighted that LLMs perform well on various natural language tasks, including sentiment analysis and the detection of slurs or derogatory remarks, demonstrating their ability to identify explicit hate and offensive speech. 

However, LLMs can be limited by their inability to process audio patterns and audio cues directly, which may result in challenges such as misidentification and false positives, especially when dealing with edge cases. Kumar et al. \cite{Kumar:2024:WYL} discovered that GPT is likely to produce false positives in identifying hate speech due to triggering on poor language (e.g., profanity, slurs) and stereotypes (34\%), even in instances of neutral or positive connotation. Given the consequences that false positives may have on online communities such as resulting in bans and suspended accounts, addressing the scalability of hate speech detection approaches in real time presents multi-faceted challenges. In addition, most previous studies focused on batch detection of harassment in text-based posts or comments \cite{Blackwell:2017:COH,Franco:2023:AUL,Jiang:2019:MCV,Jiang:2023:TCF},
a noticeable gap remains in exploring how to detect real-time verbal hate speech in VR setting that is efficient, accurate, and scalable.

To address these deficiencies and reduce false positives, in the study of \textit{Safe Guard}, we aim to bridge the gap by deploying audio feature analysis to enhance LLMs' ability to distinguish hateful content from benign content. The audio model can capture audio features such as tone, pitch, and emotion which LLMs might miss. This capability is valuable for detecting voice-based hate speech where the emotional context and speaker's intent are crucial. 

 
 We envision the future role of \textit{Safe Guard} as a first step in detecting and assisting human moderators in combating hate speech in real-time voice-based interactions. To that end, in this work we explore the first step that is concerned with the development of an LLM agent and answering the following research question (RQ): \textbf{How to detect hate speech in social VR by leveraging audio features and LLM text analysis in LLM-agents in real time?} 

The main contributions of this work include: (1) an embodied LLM-based agent to detect hate speech in real-time voice interactions in social VR, (2) a high-accuracy, fast-speed prompting method for GPT 3.5 to detect real-time voice-based hate speech, (3) a CNN classifier for extraction and analysis of audio features to assist hate speech detection, (4) a system that integrates LLM detection and audio features analysis with the LLM agent system in VRChat for real-time voice-based hate speech detection, and (5) manually collected and annotated video datasets, then transferred into audio format, for validating voice-based hate speech detection.

In the following paper, we present Literature Review, Methodology, System Design, System Evaluation Results, Discussion of the Results, Limitations, Future Work, and Conclusion.

\section{Literature Review}

\subsection{Defining Harassment and Hate Speech in the Context of Social VR}
The concept of online harassment remains highly contextual and often involves personal interpretation \cite{duggan2017online}. Prior studies pointed at the lack of consistent consensus on the definition of harassment across different online social communities. Pater et al. \cite{Pater:2016:COH} analyzed policy documents from fifteen social media platforms and pointed out that none of those platforms explicitly define what constitutes harassment. By cross-comparison of activities and behaviors that co-occur with harassment in these documents, the most common harassment types include abuse, bullying, harm, hate, stalking, and threats \cite{Pater:2016:COH}. General online harassment was also categorized by using six distinct behaviors: offensive name-calling, purposeful embarrassment, stalking, physical threats, harassment over a sustained time, and sexual harassment \cite{duggan2017online}. Alternatively, Blackwell et al. \cite{Blackwell:2017:COH} categorized toxic online harassment into four categories, namely flaming, doxing, impersonation, and public shaming. Flaming means hostile and insulting interactions between users. Doxing refers to the act of publicly revealing private information about an individual without their consent. Impersonation involves pretending to be someone else to deceive or harm, and public shaming is the act of humiliating someone in public \cite{Blackwell:2017:COH}. 

In the context of social VR, Freeman et al. \cite{Freeman:2022:DTP} examined which specific behaviors or interactions and in which context in social VR should be qualified as harassment, and concluded that any discriminatory conduct in social VR based on identity features such as gender and race (e.g., racism, misogyny, and homophobia) and with a specific target or victim are considered harassment. Moreover, Freeman et al. \cite{Freeman:2022:DTP} suggested that embodiment and immersion in social VR can simulate real-life offline physical harassment, introducing new forms of online harassment that resemble offline behaviors. Besides movement and gesture-based harassment which is unique in the social VR context, most verbal harassment on social VR platforms share similarities to those found in other online social communities, including hate speech \cite{OHagan:2023:URR}. Verbal harassment in both contexts often involves using various types of detrimental user behaviors involving abusive communications directed towards other users \cite{Zheng:2023:SRS}.

Hate speech has been defined by United Nations \cite{un_hatespeech} as ``any kind of communication in speech, writing or behaviour, that attacks or uses pejorative or discriminatory language with reference to a person or a group on the basis of who they are, in other words, based on their religion, ethnicity, nationality, race, colour, descent, gender or other identity factor.'' Guimarães, et al. \cite{Guimaraes:2023:AHS} examined well-known datasets through Web and social network crawling and 
and labeled the messages as either hate speech or non-hate speech.
By conducting experiments using four distinct datasets containing messages labeled as hate and non-hate speech, mainly from Twitter many studies such as \cite{waseem-hovy-2016-hateful, davidson2017automatedhatespeechdetection,founta2018largescalecrowdsourcingcharacterization,10.1145/3368567.3368584,10.1145/3441501.3441517} agreed on the hate speech definition as attacks on a person or a group based on race, religion, ethnic origin, sexual orientation, disability, or gender. Similarly, Arango, et al. \cite{Arango:2020:LAD} defined hate speech as communications of animosity or disparagement of an individual or a group on account of a group characteristic such as race, color, national origin, sex, disability, religion, or sexual orientation. 

\subsection{Moderation of Harassment in Social VR}
Existing research on social VR moderation emphasize the complexity of managing harassment due to its embodiment nature, identity-related threats, presence of minors, and lack of consensus on definitions \cite{Zheng:2023:SRS,Fiani:2024:Perspectives}. The immersive and real-time nature of voice-based interactions in social VR makes it difficult for social VR platforms to accurately and effectively detect harassment due to the absence of a persistent, written record \cite{Jiang:2019:MCV}. Bad actors can deliver insults and abusive comments in real time and cause immediate and direct harm. Schulenberg et al. \cite{Schulenberg:2023:AI_Moderation} claimed that social VR might lead to more severe forms of harassment compared to other contexts. Therefore, hate speech misconducts such as racial slurs in voice channels faced harsher punishments than in text channels \cite{Jiang:2019:MCV}.


In the context of social VR, Schulenberg et al. \cite{Schulenberg:2023:AI_Moderation} provided an overview of moderation efforts including Community Guidelines and Punishments, and Moderation Pipelines that are mainly based on human moderation, and introduced the potential of AI moderation in social VR. They concluded that the awareness of the presence of an AI moderation system that can detect the large-scale embodied and immersive multi-user virtual environments and act in the moment can effectively prevent harassment in social VR before it occurs \cite{Schulenberg:2023:AI_Moderation}. Existing safety-enhancing features on social VR platforms such as blocking, personal space bubble, muting, reporting players or trust systems \cite{Fiani:2024:Perspectives} to keep users safe from problematic users were found to have limitations \cite{Maloney:2020:IC,Fiani:2024:PEM}. First, they place the burden of moderation on users, who might be ill-equipped or unaware of effective strategies \cite{Hartikainen:2016:DCI}. Second, platforms rely on volunteer moderators to manage behavior in public virtual rooms \cite{Blackwell:2019:HSV}, but they cannot cover most incidents. Sabri et al. \cite{Sabri:2023:CMS} showed that only 24\% of incidents in social VR are addressed by moderators, highlighting the need for better moderation tools.

Automated embodied moderation has been investigated as an alternative solution \cite{Fiani:2024:PEM} to create safer spaces in social VR by providing a protective figure that takes action to mitigate harmful interactions. Previous studies \cite{Fiani:2024:PEM,Fiani:2023:BB} explored the use of VR-embodied AI agents to maintain safe and respectful interactions in social VR. Fiani et al. \cite{Fiani:2023:BB} introduced “Big Buddy” as a Wizard-of-Oz automated agent in a simulated social VR game and intervenes when a fictional player disrupts the child’s game. Unlike traditional moderation tools(e.g., reporting and blocking) that operate in the background and would have flaws in the process and lack trust and feelings of unfair treatment discouraging users from using them \cite{Fiani:2023:BB}, embodied agents are visually present within the virtual environment, allowing for real-time interaction and enforcement of community guidelines. However, these AI-based moderation systems have key limitations in their ability to detect problematic events, parse ambiguity, and identify false positives \cite{Fiani:2023:BB,Lai:2022:HAC}. False positives are produced by mistakenly flagging benign content as harmful \cite{Kolla:2024:LLM}. Misidentifying content as harmful or inappropriate can lead to unjust removals or sanctions, which could harm users who may be unfairly targeted \cite{rahaman2024}.

Dealing with uncertainty and missing contextual factors (e.g., tones and pitch in voice) remains a key challenge for automated embodied moderation \cite{Fiani:2024:PEM}. Recent advances in AI, particularly in LLMs, along with immersive VR and avatar interfaces, allow for the creation of embodied conversational agents that can engage in natural dialogue with humans \cite{Paudel:2024:CSM}. As multiple experimentations \cite{csepregi2023,vanStegeren:2021:FGD,Ashby:2023:QDG,Ammanabrolu:2021:DG} have explored incorporating LLMs into LLM-agent conversations, LLM-based agents are becoming more and more prevalent. LLMs can quickly generate dynamic and high-quality responses that adjust to a player’s actions, choices, and the overall state of the game world \cite{csepregi2023,Paudel:2024:CSM}. Mehta et al. \cite{Mehta:2022:CAI} highlighted that conversational AIs and LLMs in LLM-agent interactions as 'extremely viable' and a potential successor of traditional pre-scripted dialogues. Therefore, the emergence of VR-embodied LLM-based agents in social VR opens up underexplored opportunities to leverage LLM's potential for not only generating conversations but also capturing and detecting real-time voice-based hate speech. 


\subsection{LLMs for RT Voice-based Hate Speech Detection}
LLMs have shown promise in real-time (RT) voice-based hate speech detection due to their capability for identifying the violating content and providing the reasoning on why the content violated rules \cite{Kolla:2024:LLM}. LLMs showed good contextual understanding, which allows them to provide correct reasoning when handling rule violations in online communities \cite {Kolla:2024:LLM}. LLMs also have demonstrated state-of-the-art performance in natural language tasks. LLMs have undergone extensive training using vast amounts of natural language data \cite{guo2024investigation}, enabling them to grasp intricate contextual details. This feature enables LLMs to detect RT verbal hate speech in complex or ambiguous scenarios, where traditional approaches fall short for the existing supervised learning-based detection methods cannot fully capture context to make accurate predictions \cite{guo2024investigation}.

The robustness of LLMs to understand and interpret text right after conversion from audio, even when it contains typographical errors, slang, or informal language may contribute to faster detection times in the real-time hate speech detection. Traditional machine learning models and deep learning approaches require extensive text preprocessing to normalize the input data, including tokenization, stop words removal, stemming, and lemmatization \cite{Kadhim:2018:EPT}. These steps are essential for ensuring that the input data is consistent and interpretable. On the contrary, LLMs can comprehend the meaning and context of raw text inputs without requiring preprocessing owing to their extensive pre-training, which could translate into a more streamlined and efficient pipeline.


Additionally, LLMs require minimal feature engineering, streamlining the development process and allowing for faster deployment. This is particularly advantageous in real-time voice moderation, where rapid and accurate detection is important for maintaining a safe and inclusive social VR environment. In comparison, traditional and deep learning Machine Learning methods such as Recurrent Neural Networks, Deep Neural networks, and long short-term memory networks require extensive feature engineering and large labeled datasets \cite{Grondahl:2018:AYN}. Fine-tune pre-trained transformers such as BERT, RoBERTa and ALBERT \cite{Rana2022EmotionBH} are resource-intensive during both training and inference.


\textit{Limitations of LLMs in Hate Speech Detection}
LLMs are not able to detect tones and emotions within verbal interactions, particularly in edge cases where tone and emotions are crucial for differentiating the nature of verbal speech \cite{Kumar:2024:WYL}. Rana et al. \cite{Rana2022EmotionBH} stated that the most essential features in classifying hate speech would be the speaker’s emotional state and its influence on the spoken words. Tones, emotions, and vocal intonations play crucial roles in conveying the intent behind words. The same sentence can have different meanings based on the speaker's tone, whether to be playful, sarcastic, angry, calm. For example, Rana et al. \cite{Rana2022EmotionBH} asserted that a political leader calmly discussing immigration policies at a conference is less harmful than delivering the same speech with extreme anger and disgust towards a specific targeted user, as the latter incites hostility against immigrants in the country. Another example is the different interpretations of certain words depending on the emotion and context \cite{Rana2022EmotionBH}. When a friend jokingly calls someone a name in a playful tone and the conversation remains calm or civil, it should not be classified as hate speech. However, if the same words were delivered with a harsh tone and intent to attack specific individuals, they should be detected as offensive or hurtful and classified as hate speech. Since LLMs rely solely on plain text, they are unable to capture the subtleties of spoken language. This limitation results in misclassifications and inaccuracies, especially false positives, of verbal harassment in hate speech detection. False positives often due to LLMs' difficulty in accurately interpreting human emotions \cite{Kumar:2024:WYL}, as important contextual cues present in tone are not conveyed through text. For example, LLM-based moderator might take the user’s exaggerated
language as disrespectful to others \cite{Kumar:2024:WYL}. 

\subsection{Improving the Accuracy of LLM Using Audio Feature Analysis for Hate Speech Detection}
Audio features can be analyzed and applied to address the shortcomings of LLMs in detecting hate speech. Rana et al. \cite{Rana2022EmotionBH} suggested that the classification of hate speech is significantly influenced by the speaker’s emotional state and its impact on their spoken words. The study claimed that incorporating emotion into hate speech detection can help reduce false positives in systems that rely solely on text data as input. Patrick et al. \cite{Patrick1901ThePO} also asserted that there is a close link between hate speech and the emotional and psychological state of the speaker and evident in the emotional tone of their language. 

Kumar et al. \cite{Kumar:2024:WYL} showed that incorporating context in LLM rule-based moderation corrected 35\% of errors with minimal changes to the prompt. Integrating audio feature analysis with LLMs can enhance hate speech detection by providing context that text alone cannot capture. Barakat et al. \cite{Barakat:2012:DOV} demonstrated that MFCC audio features significantly improved the accuracy of independent keyword spotting (KWS) for detecting offensive language in video blogs, greatly outperforming the existing speech-to-text methods, which indicates the potential to use audio features for detecting audio-based hate speech. 

Das et al. \cite{HateMMarticle} identified that hate speech demonstrated a certain pattern in temporal features including zero crossing rate and root mean square energy. This pattern can be used to assist hate speech detection by capturing distinctive acoustic characteristics. Analyzing elements like pitch, volume, and speech rate gives insights into the speaker's emotional state and intent, thereby improving classification accuracy. Khan et al. \cite{Khan:2017:ERU} highlighted the effectiveness of combining audio features such as pitch and Mel-Frequency Cepstral Coefficients (MFCC) with models like K-Nearest Neighbors (KNN) and Naive Bayes to enhance emotion classification accuracy. Hu et al. \cite{Hu:2007:GMM} found that using MFCC features with a Gaussian Mixture Model (GMM) combined with a Support Vector Machine (SVM) outperformed traditional GMM methods, achieving improvements in emotion classification accuracy. Similarly, Zhou et al. \cite{Zhou:2009:SER} introduced GMM supervector with SVM using MFCC features achieving an 88\% accuracy rate in classifying emotions.


Finally, recent work by Roblox \cite{blog_key} highlights the advancement of combining a proprietary text filter and audio features analysis for detecting hate speech. They demonstrated the efficacy of a CNN model to create classifiers on MFCC audio features. The model can detect and categorize verbal hate speech into different categories including profanity, dating and sexual, racism, and bullying. This combined method achieved an average precision of 94.48\% \cite{blog_key}, demonstrating the effectiveness of integrating audio features with text analysis to enhance detection capacity. We are extending this work by leveraging publicly available GPT model and aim at reducing false positives.

\section{Methodology}
Our approach involves the following key steps: 
\begin{itemize}
    \item Defining voice-based hate speech criteria in social VR,
    \item Designing and implementing an LLM-based agent \textit{Safe Guard} for hate speech detection,
    \item Exploring suitable LLMs prompting method for real-time hate speech detection,
    \item Processing voice messages through text conversion and applying real-time LLM-based detection on transcripts. In parallel, extract and analyze the key features of audio using a CNN classifier to assist detection, 
    \item Building and integrating the LLM-based detection system into the \textit{Safe Guard} agent architecture to enhance moderation capabilities,
    \item Collecting and analyzing evaluation metrics to iteratively refine and optimize the proposed system.
\end{itemize}

\subsection{Hate Speech Training and Testing Datasets}
To the best of our knowledge, there are no publicly available audio datasets for hate speech detection. To meet the purpose of our real-time voice-based hate speech detection, we used the HATEMM dataset \cite{HateMMarticle}, which was sourced from the BitChute platform and consists of 1083 videos annotated for hate speech. This dataset includes a ground truth annotation file, with 39.8\% of the samples labeled as hate speech. We extracted audio from these videos and transcribed it into text using the OpenAI Whisper. The dataset was split into 80\% for training purposes, and 20\% for testing. The processing approaches involved extracting audio from the videos using FFmpeg API. The audio was then converted into text transcripts with OpenAI Whisper API, and the transcripts were stored in files. Further processing was conducted by combining the transcripts with extracted audio features to improve detection accuracy.

\subsection{LLM Set Up and Prompt with Hate Speech Moderation Rules}
In this study, we employed ChatGPT 3.5 as our LLM model for hate speech detection in social VR. Our choice of this model was due to several key considerations. Firstly, GPT 3.5 is capable of handling complex contextual nuances. Secondly, GPT 3.5 is more practical in real-time detection because it requires less computational resources and is more efficient than GPT 4. Overall, it offers a good balance between performance and resource usage.

Before using LLMs for real-time detection, we provided information to the models on datasets labeled for hate speech. Prompt-based strategies have been found to effectively guide LLMs in leveraging the context of the specific task \cite{10.1145/3560815}. A prompt is a query or statement designed to instruct the model on what is being asked. The effectiveness of an LLM can be significantly improved with carefully crafted prompts, underscoring the importance of prompting techniques in leveraging the context of these models. Guo, et al. \cite{guo2024investigation} claimed that the effectiveness of LLMs in identifying hate speech is highly contingent upon the design of the prompt. 

\subsection{Convolutional Neural Network Audio Feature Model}
The CNN model is a class of deep learning algorithms primarily used for processing structured grid data, which is suitable for our audio parameters. CNN model is capable of using backpropagation through different layers to build an accurate classifier. Convolutional layers use small learnable filters to identify features across the input data. Activation layers like ReLU introduce non-linearity by converting negative values to zero, assisting the network in learning patterns. Pooling layers reduce the spatial dimensions of the data and reduce the computational load. Following several convolutional and pooling layers, the data is flattened into a 1D vector to classify at last.

\section{System Design}
\subsection{Safe Guard Agent Design}
 The embodied LLM agent \textit{Safe Guard} is designed to facilitate real-time interactions with players while proactively monitoring and moderating conversations in VRChat. Building on the previous work by Park et al. \cite{park2023generativeagentsinteractivesimulacra} and Wan et al. \cite{10.1145/3613905.3651026}, \textit{Safe Guard} incorporates several features for context management and moderation. The system maintains memories of interactions, capturing not only the dialogue but also the player's attitude, inferred mood based on input messages, and the time when the conversation took place \cite{10.1145/3613905.3651026}. This comprehensive memory is crucial for understanding and interpreting ongoing interactions in social VR. To manage and evaluate context effectively, \textit{Safe Guard} uses advanced techniques such as exponential decay to prioritize recent interactions and cosine similarity metrics to assess the relevance of past conversations \cite{10.1145/3613905.3651026}. These methods ensure that \textit{Safe Guard} accurately evaluates each interaction's importance and relevance to current discussions. 

\textbf{Dual Mode Operation: } \textit{Safe Guard} can operate in both conversational mode (engaging with a single user) and observational mode (monitoring group interactions), switching as needed based on the scenario.

Key aspects of the agent system are defined below:

\textbf{\textit{GPT Module.}}
ChatGPT is deployed throughout the process of \textit{Safe Guard} system design to process user input, understand the context of social situations and generate appropriate observations or responses. It not only facilitates human-like interactions but also actively monitors for potential instances of hate speech. 

\textbf{\textit{LLM-Agent Formation Module.}}
LLM-agent is created by assigning it a base description that includes name, character details, preferences, etc \cite{10.1145/3613905.3651026}. \textit{Safe Guard} is tailored specifically for hate speech detection. Its character details are "\textit{A vigilant, neutral, and approachable guardian, focused on maintaining respectful communication within VR spaces}". Unity engine was used to modify the chosen avatar to include a set of animations for expressions and actions. 

\textit{\textbf{Observation Database Module.}}
The system maintains a collection of context observations that store all observations generated by LLM for each conversation. This includes initial observations from the LLM-agent’s "Base Description". Further conversational contexts are captured by generating up to three observations for each message \cite{10.1145/3613905.3651026}. Memory observations are defined as combinations of Base descriptions and Context observations (see \cite{10.1145/3613905.3651026} for more details on the LLM-agent itself).

\begin{figure}
    \centering
    \includegraphics[width=1\linewidth]{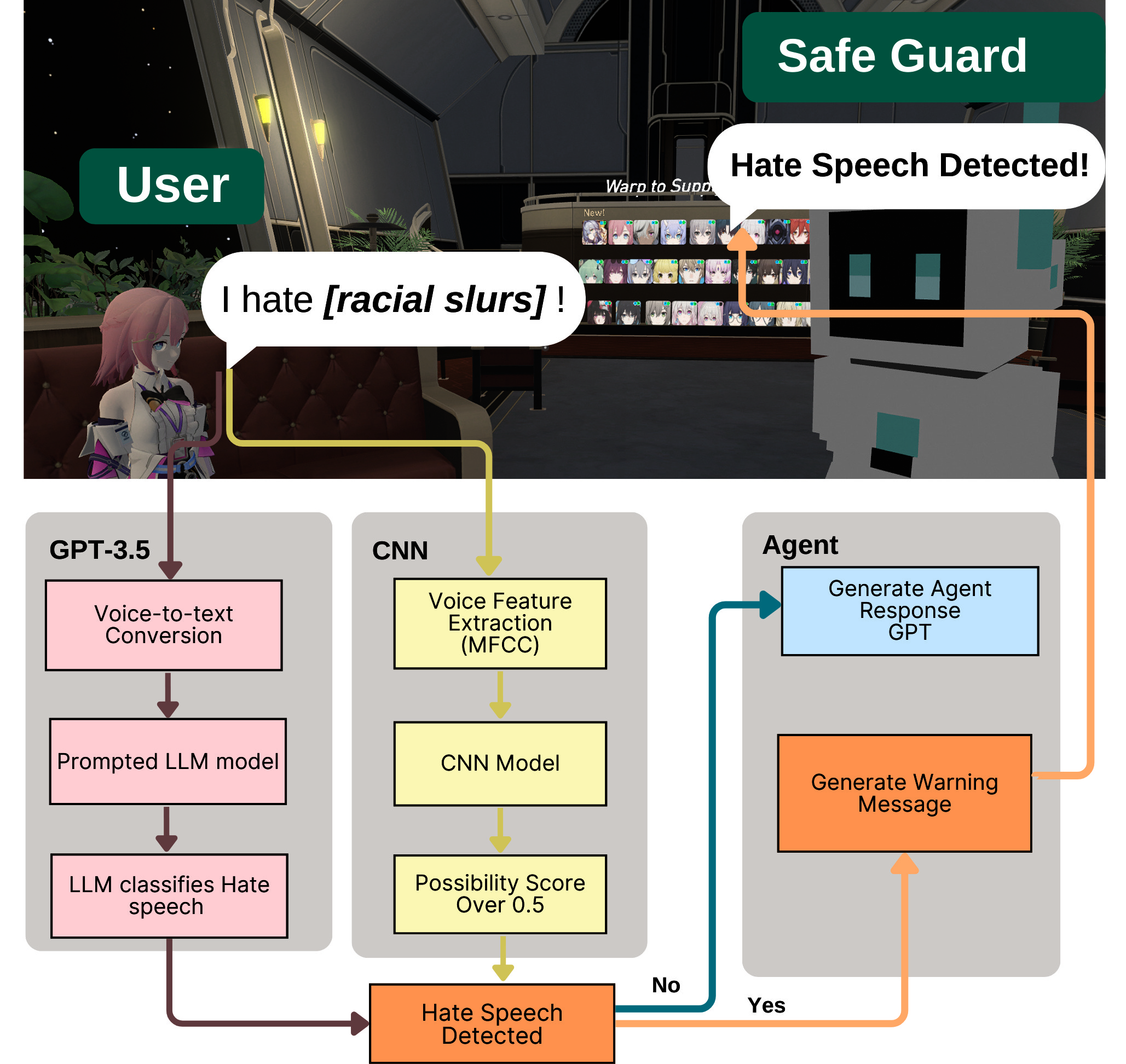}
    \caption{Safe Guard System Design in Conversational Mode}
    \label{fig: SystemDesign}
\end{figure}

\subsection{Hate Speech Detection System}
The real-time hate speech detection system is designed to integrate into VRChat, ensuring a safe and respectful environment. As shown in \cref{fig: SystemDesign}, the flowchart illustrates the workflow of the proposed \textit{Safe Guard} hate speech detection system in conversational mode. The system operates similarly in observational mode when multiple players are presented, except that the LLM-agent does not generate dialogue with the players. 

\subsection{Prompt Engineering for LLM Model}
To set up our hate speech detection system, we used specific prompts for ChatGPT 3.5 based on established hate speech definitions and reasoning to produce a binary response. We compared and decided on the most suitable and effective prompting method in the experiment stage. Since the project is applied to real-time detection, we focused on prompt methods that do not need lengthy context or multiple sequential prompts. We mainly evaluated three different approaches as discussed below.

\subsubsection{Approach \#1: Direct Prompting}
For this method, we used a straightforward prompt to ask GPT to classify text as either "Hate" or "Non Hate" without providing any extra instruction. This prompting method has been proven to be effective in previous study \cite{10.1145/3643829}. Using the general prompt method can rapidly enable the LLM model to complete the voice hate speech detection task. The prompting message we used for this method was: \textit{"You are a classifier that identifies whether a given text contains hate speech or not. Is the following text hate speech? \
Respond with Yes or No. Here is the text:     "}

\subsubsection{Approach \#2: Prompting with Hate Speech Definition}
This method involved providing a clear definition of hate speech within our use case and asking if a specific text fits that definition. The prompting message we used for this method was: \textit{"Hate speech in social VR includes any speech that attacks a person or group based on attributes such as race, religion, ethnic origin, national origin, sex, disability, sexual orientation, or gender identity. Consider this context carefully. Is the following text hate speech? \
Respond with Yes or No. Text:     "}

\subsubsection{Approach \#3: Few-Shot Learning Prompting}
Besides the definition of hate speech, we provided several examples with ground truth labels to help GPT learn what constitutes hate speech and what does not. The prompting message we used for this method was: \textit{" Hate speech includes any speech that attacks a person or group based on attributes such as race, religion, ethnic origin, national origin, sex, disability, sexual orientation, or gender identity. Consider this context carefully.}

\textit{Here are some examples of hate speech and non-hate speech:}

\textit{Example 1:
Text: "I think people of all races should be treated equally."
Hate Speech: No}

\textit{Example 2:
Text: "People from [name] country are all criminals and should be deported."
Hate Speech: Yes}

\textit{Example 3:
Text: "Everyone deserves to be loved, regardless of their gender identity."
Hate Speech: No}

\textit{Example 4:
Text: "I think all immigrants should get out of Canada."
Hate Speech: Yes}

\textit{Determine if the following text is hate speech. Respond with Yes or No.
Text:       "}



\subsection{Audio feature Extraction}
Voice consists of voice signals and their content. Apart from content, voice signals can also provide information for hate speech detection.  Feature extraction is the process of extracting and tackling hidden information in the raw data signal \cite{Ali:2021:MAR}. We used root mean square(RMS) and Mel-Frequency Cepstrum Coefficients(MFCC) for audio feature analysis.

\textbf{Root mean square (RMS)} is the square root value of the mean of the sum of squares of the signal. In the context of audio signals, RMS is often used to measure the power or loudness of the signal \cite{HateMMarticle}. It provides a single value that represents the average energy of the waveform, which is particularly useful for analyzing the amplitude of audio signals over time. 
\begin{equation}
\text{RMS} = \sqrt{\frac{1}{N} \sum_{i=1}^{N} x[i]^2}
\end{equation}

where:
\begin{itemize}
  \item $x[i]$ is the amplitude of the signal at the $i$-th sample,
  \item $N$ is the total number of samples.
\end{itemize}

\textbf{Mel Frequency Cepstrum Coefficients (MFCCs)} are widely used in speech and audio processing as a representation of the short-term power spectrum of sound \cite{Majeed:2015:MFCC,Ali:2021:MAR}. They are useful in various tasks such as speech recognition, speaker identification, emotion recognition, and audio classification. Given their utility, incorporating MFCCs into our hate speech detection method is beneficial. Ali et al. \cite{Ali:2021:MAR} stated that MFCC can be calculated by
conducting five consecutive processes, namely signal framing, computing of the power spectrum, applying a Mel filter
bank to the obtained power spectra, calculating the logarithm
values of all filter banks, and finally applying the Discrete Cosine transform (DCT). 

After an initial exploration of audio features, we decided to use the 40-dimensional vector MFCC features in our CNN model, in order to capture key features of the user input audio data. To extract these features from WAV audio files, we used the Python library librosa and stored the results in a CSV file for further analysis.



\subsection{Audio Feature Model}
After extracting features of audio files, we used machine learning techniques to build a CNN classifier based on those features. 


Initially, we set up a sequential model consisting of a linear stack of layers. The input layer included a 2D convolutional layer with 32 filters, each of size 3x1, designed to extract spatial features from the input data, which has an input shape of 40 dimensions with 1 feature. Next, we added a second convolutional layer and a second max pooling layer. Finally, a flattening layer was used to convert the 2D data into 1D, followed by two fully connected dense layers to perform the final classification.

\subsection{Using a Combination of LLM and Audio Feature Model to Detect Hate Speech}

In our proposed system, we combined GPT 3.5 as the primary method with an audio-based CNN classifier to enhance the overall detection performance. The integration of these two methods aims to combine the strengths of both NLP and audio signal analysis to improve the accuracy and reliability of the detection results.

After converting audio to text, the transcript is delivered to the prompted LLM model. The prompted LLM model goes through the transcript and detects whether hate speech is in it. Simultaneously, audio key features are extracted and analyzed by the CNN classifier to generate a probability score. If this probability score is greater than 0.5, the input is classified as 'Hate.' A threshold of 0.5 is commonly used for making binary classification decisions.

To determine the final classification result of the combined model, we applied the following decision rule: The audio input is classified as 'hate speech' only if both the GPT model and the audio-based CNN classifier predict it as 'hate.' Otherwise, the final classification is 'non-hate speech.' Once the system successfully detects hate speech, the system will immediately display a voice notification in the agent's dialogue to indicate that hate speech has been detected (see Figure \ref{fig: SystemDesign}). 

\subsection{Integration to Safe Guard Agent System}
We integrated the hate speech detection system into the \textit{Safe Guard} LLM-agent and processed the audio within the module to determine if hate speech was present. Once hate speech is detected, the program will immediately display a warning message in the \textit{Safe Guard} dialogue. We fine-tuned the silent detection parameters, including increasing the max silence length to 2 seconds and lowering the threshold to -40db for activating the recording procedure. In this way, we were able to retrieve segmented sentences instead of a whole speech for audio processing. We also altered the sample rate to 44100Hz, which is the maximum rate that is compatible with the \textit{Safe Guard} agent system to get better recording quality and support audio feature extraction. 

\section{Experiment}
\subsection{Procedure}
\textit{VRChat System Setup for Experiment}: To simulate the real scenario environment in VRChat, we connected two devices that log in with different accounts in VRChat.One account operated as the host server, running the LLM agent and detection system to capture audio from the test subject. Meanwhile, the second account logged in as the player, transmitting the test speech dataset via microphone. 

During the \textit{audio feature process}, the combination of RMS and MFCCs played a key role in classifying the hate and non-hate datasets, yielding the best results for feature selection. Extracted features went through a CNN training model, which produced a similarity score to the recognized hate speech pattern. 

\textit{Prompting Experiment}: We evaluated the effectiveness of our LLM hate speech detection system using different prompting techniques. We tested with the 50 videos containing hate speech from our primary MM Hate dataset. True labels for each data will be derived from the manual categorization of the video dataset.
 

\subsection{Evaluation Metrics}
\subsubsection{Latency: Real-Time Detection Time}
To evaluate the performance of the real-time hate speech detection system, we measured the latency at each stage of the detection process. The overall latency includes four main components as follows: (1) \textbf{Audio Feature Extract Time}: The duration of the audio feature extraction from raw audio data, (2) \textbf{Speech-to-Text Conversion Time:} The time taken to transcribe the audio input into text using a speech recognition engine,
(3) \textbf{Audio Feature Prediction Time:} The time taken by the CNN classifier to classify the audio based on audio features as hate speech or non-hate speech, and (4) \textbf{ LLM Analysis Time:} The time taken by the LLM to classify the text as hate speech or non-hate speech.

The total detection time is the sum of these individual times, representing the overall latency from receiving the audio input to providing the moderation output.

\subsubsection{Accuracy Metrics}
To evaluate the accuracy of our real-time voice-based hate speech detection system in social VR, we utilized several key metrics that are well-known and academically validated \cite{guo2024investigation} as follows:

\textbf{Accuracy:} Accuracy measures the proportion of correct detections (both true positives and true negatives) out of all detections. It is given by:
\[
\text{Accuracy} = \frac{\text{True Positives (TP)} + \text{True Negatives (TN)}}{\text{Total Instances}}
\]

\textbf{Precision:} Precision measures the proportion of true positive detections (correctly identified instances of hate speech) out of all positive detections (instances flagged as hate speech). It is calculated as:
\[
\text{Precision} = \frac{\text{True Positives (TP)}}{\text{True Positives (TP)} + \text{False Positives (FP)}}
\]

\textbf{Recall (Sensitivity):} Recall evaluates the proportion of true positive detections out of all actual instances of hate speech. It is defined as:
\[
\text{Recall} = \frac{\text{True Positives (TP)}}{\text{True Positives (TP)} + \text{False Negatives (FN)}}
\]

\textbf{F1 Score:} The F1 score is the harmonic mean of precision and recall, providing a single metric that balances both. It is calculated as:
\[
\text{F1 Score} = 2 \times \frac{\text{Precision} \times \text{Recall}}{\text{Precision} + \text{Recall}}
\]
\section{RESULTS}





\subsection{Validation Dataset}
A total of 204 video clips were collected using search API from YouTube with manually annotated true labels for validation. 104 of them are hate speech, and 100 of them are non-hate speech. We filtered the data within 20 seconds to better simulate real conversation scenarios in the VRChat setting.

\subsection{Data Analysis}
\subsubsection{GPT3.5 Prompting Methods Comparison Analysis}

Table.~\ref{tab:gpt_performance_comparison} below compares solely the GPT's performance across three prompting methods discussed above for hate speech detection using our training dataset. Since the datasets consist of long narrative speech rather than short conversational dialogues, the results did not show great accuracy. 

\begin{table}[h!]
\centering
\begin{tabular}{|c|c|c|c|c|} \hline 

\textbf{Method} & \textbf{Acc. (\%)} & \textbf{Prec. (\%)} & \textbf{Rec. (\%)} & \textbf{F1 (\%)} \\ \hline  
\multicolumn{5}{|c|}{\textbf{Direct}} \\ \hline  
Non Hate &  & 71.43 & 19 & 29 \\ \hline  
Hate &  & 56 & 93 & 70 \\ \hline  
Overall & 57.52 & 63 & 56 & 50 \\ \hline  
\multicolumn{5}{|c|}{\textbf{Definition Prompt}} \\ \hline  
Non Hate &  & 67 & 77 & 71 \\ \hline  
Hate &  & 91 & 85 & 88 \\ \hline  
Overall & 82.98 & 79 & 81 & 80 \\ \hline  
\multicolumn{5}{|c|}{\textbf{Few-shot}} \\ \hline  
Non Hate &  & 78 & 78 & 78 \\ \hline  
Hate &  & 93 & 93 & 93 \\ \hline  
Overall & 85 & 89 & 89 & 80 \\ \hline 
\end{tabular}
\caption{Comparison of GPT-3.5 Performance for Hate Speech Detection}
\label{tab:gpt_performance_comparison}
\end{table}

Among the three methods, few-shot prompting emerged as the most effective method for hate speech detection, outperforming the other two methods. It achieved higher accuracy and balanced precision and recall, making it a more suitable choice in our use case.



\subsubsection{Detection System Result Analysis}

\textbf{Comparison with Baseline Models:}
We compared the performance metrics of our proposed combination model against the GPT-3.5 alone, and the audio feature model alone as two baseline models. The results are shown in Table.~\ref{tab:classification-report} which displays the classification performance evaluation report including Precision, Recall, and F1-score for both hate and non-hate categories. 

\begin{table}[h!]
\centering
\begin{tabular}{|l|l|c|c|c|}
\hline
\textbf{Model} & \textbf{Class} & \textbf{Precision} & \textbf{Recall} & \textbf{F1-Score} \\ \hline
\multirow{5}{*}{GPT} 
& Hate & 0.95 & 0.95 & 0.95 \\ \cline{2-5} 
& Non-Hate & 0.94 & 0.94 & 0.95 \\ \cline{2-5} 
& \multicolumn{2}{|l|}{Accuracy} & \multicolumn{2}{c|}{0.95} \\ \cline{2-5} 
& Macro avg & 0.95 & 0.95 & 0.95  \\ \cline{2-5} 
& Weighted avg & 0.95 & 0.95 & 0.95  \\ \hline
\multirow{5}{*}{Audio} 
& Hate & 0.73 & 0.53 & 0.62 \\ \cline{2-5} 
& Non-Hate & 0.63 & 0.79 & 0.71 \\ \cline{2-5} 
& \multicolumn{2}{|l|}{Accuracy} & \multicolumn{2}{c|}{0.67} \\ \cline{2-5} 
& Macro avg & 0.68 & 0.67 & 0.66 \\ \cline{2-5} 
& Weighted avg & 0.68 & 0.67 & 0.66 \\ \hline
\multirow{5}{*}{Combined} 
& Hate & 0.96 & 0.53 & 0.69 \\ \cline{2-5} 
& Non-Hate & 0.67 & 0.99 & 0.80 \\ \cline{2-5} 
& \multicolumn{2}{|l|}{Accuracy} & \multicolumn{2}{c|}{0.75} \\ \cline{2-5} 
& Macro avg & 0.82 & 0.76 & 0.74 \\ \cline{2-5} 
& Weighted avg & 0.82 & 0.76 & 0.74 \\ \hline
\end{tabular}
\caption{Classification Report for GPT Model, Audio Model, and Combined Method}
\label{tab:classification-report}
\end{table}

\begin{figure*}[!htb]
    \centering
    \begin{minipage}{0.32\linewidth}
        \centering
        \includegraphics[height=5cm]{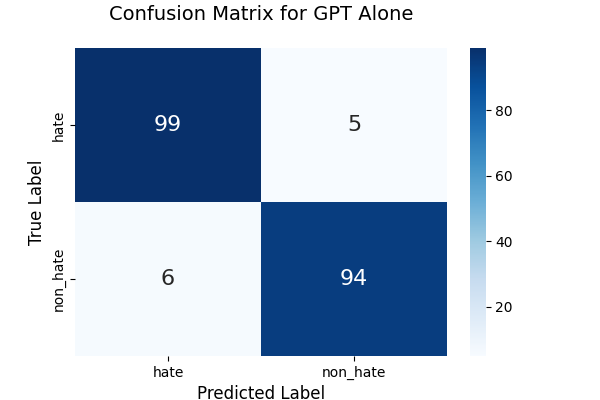}
        \caption{GPT Model Alone}
        \label{fig: gpt_alone}
    \end{minipage}\hfill
    \begin{minipage}{0.32\linewidth}
        \centering
        \includegraphics[height=5cm]{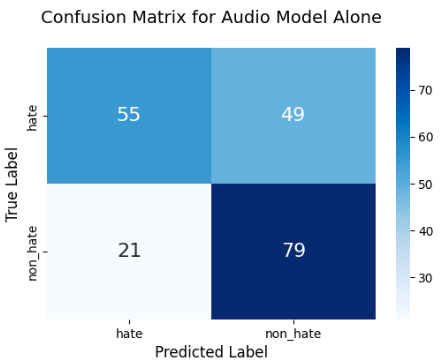}
        \caption{Audio Feature Model Alone}
        \label{fig: audio_alone}
    \end{minipage}\hfill
    \begin{minipage}{0.32\linewidth}
        \centering
        \includegraphics[height=5cm]{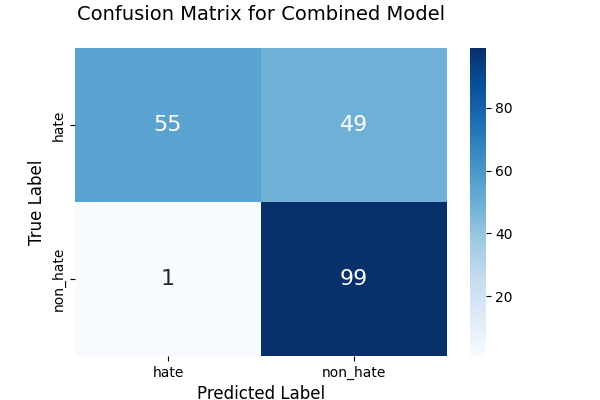}
        \caption{Combined Model}
        \label{fig: combined}
    \end{minipage}
\end{figure*}

 The GPT Model showed strong and consistent performance across all metrics (Fig.~\ref{fig: gpt_alone}). For both hate and non-hate speech, it achieved high Precision and Recall of around 0.95. The high accuracy indicated that the GPT Model was very reliable for detecting both hate and non-hate speech. However, the downside was that the false positive rate was fairly high compared with the combined model.

The Audio Model exhibited lower performance than GPT model (Fig.~\ref{fig: audio_alone}). For hate speech, it recorded a Precision of 0.73 and a Recall of 0.53. This indicated that the model struggled with false negatives in hate speech detection. In the case of non-hate speech, the model performed slightly better, with a Precision of 0.63, and a Recall of 0.79. Overall, the Audio Model achieved an accuracy of 0.67, reflecting moderate effectiveness.

The combined Method exhibited a more balanced performance (Fig.~\ref{fig: combined}). For hate speech, it achieved a high Precision of 0.96 but a lower Recall of 0.53, resulting in an F1-Score of 0.69. For non-hate speech, the Combined Method showed a Precision of 0.67 and an outstanding Recall of 0.99. The overall accuracy of 0.75 indicated that the Combined Method outperforms the Audio Model, and can enhance the GPT Model detection results.\\

\textbf{Comparison of False Positive Rate (FPR) across 3 Models:} FPR is calculated as FP / FP+TN, where FP is the number of false positives and TN is the number of true negatives. As the results are shown in Fig.~\ref{fig: gpt_alone}, GPT model's FPR was 0.06, indicating that 6\% of the non-hate speech cases were mistakenly classified as hate speech. Meanwhile, the Audio Feature Model exhibited a much higher FPR of 0.21 (Fig.~\ref{fig: audio_alone}), exhibiting a greater tendency to misclassify non-hate speech. In contrast, the Combined Model performed significantly better, with an improved FPR of just 0.01 (Fig.~\ref{fig: combined}), meaning it only misclassified 1\% of the non-hate speech cases as hate speech, which illustrated the effectiveness of the combined approach in reducing false positives compared to the individual models. 

\subsubsection{Latency Measurement} 
We collected and analyzed latency measurements during different stages of the pipeline. We identified the average, minimum, and maximum latency times during tests. \cref{tab: processing_times} shows the processing times for different stages of the system.

\begin{table}[ht]
\centering
\begin{tabular}{|l|c|c|c|}
\hline
\textbf{Process} & \textbf{Min (s)} & \textbf{Mean (s)} & \textbf{Max (s)} \\ \hline
Transcription     & 0.48   & 0.95   & 1.99   \\ \hline
LLM Analysis      & 0.27   & 0.48   & 1.83   \\ \hline
Audio Extraction  & 0.006  & 0.019  & 0.045  \\ \hline
Audio Prediction  & 0.05   & 0.055  & 0.065  \\ \hline
\end{tabular}
\caption{Processing times for different stages of the system.}
\label{tab: processing_times}
\end{table}

\begin{figure}[h]
    \centering
    \includegraphics[width=1\linewidth]{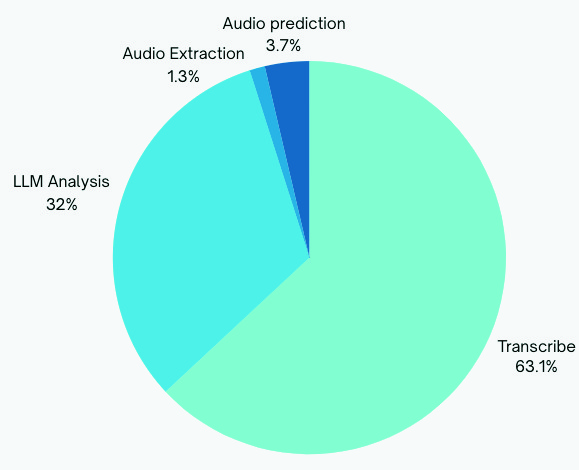}
    \caption{Overall Latency Distribution}
    \label{fig: pie chart}
\end{figure}

Our tests demonstrated that the average latency for our proposed combined detection system is approximately \textbf{1.5 seconds}, which we considered to be acceptable for our real-time verbal hate speech detection use case. The chart in Fig.~\ref{fig: pie chart} illustrates that on average over 60\% of the overall latency is due to transcribing, and 32\% is attributed to GPT processing. Audio feature extraction and prediction time only accounted for around 5\%. The findings align with our earlier discussion of the pros of cons of LLMs and audio feature analysis in the previous sections.

\section{DISCUSSION}
To answer our RQ we leverage the power of LLMs and audio features to address the complex, often ambiguous nature of hate speech detection in real-time, interactive environments in social VR.


\textbf{Effectiveness and Trade-off of LLM for Hate Speech Detection} 
While the GPT Model demonstrated great overall performance with a high Accuracy of 0.95, it also exhibited a trade-off of false positives. The GPT Model's Precision and Recall for hate speech detection were both strong, but its tendency to misidentify non-hate speech as hate speech led to a higher false positive rate. This may result in unnecessary moderation requests and potential user experience issues \cite{Lai:2022:HAC}.

\textbf{Integration of Audio Features with LLMs to Reduce False Positives}
Incorporating audio features into GPT presented a more balanced approach to hate speech detection. Although the overall Accuracy of the Combined Method was lower at 0.75 compared to the GPT Model alone, it demonstrates advantages in minimizing false positives. The Combined Method also achieved a high Precision of 0.96 for hate speech detection, indicating that when it identified hate speech, it was highly accurate. This high Precision meant fewer cases of non-hate speech being incorrectly labeled as hate speech, which can enhance user experience by reducing the likelihood of wrongful classifications and aligns better with real-world business needs. 


By ensuring the detected hate speech was truly hate, the combined method can reduce the risk of incorrectly identifying non-hate speech as hate speech, which might harm the users' experience. Additionally, the high accuracy of the true positive hate speech cases also showed the potential of the Combined Method to build an automated hate speech moderation pipeline, which could reduce the need for human moderation involvement. 


\textbf{Overcoming Misclassification Caused by Audio Interference}
In our review of misclassified cases, we identified several key factors contributing to detection errors, including poor audio quality, background noise (music or singing), and instances where the speaker was laughing while speaking. These issues made it difficult for both the audio model and GPT to interpret speech accurately. Addressing these challenges is crucial for enhancing the accuracy of real-time hate speech detection in social VR. A possible solution to these challenges is drawn from the previous study \cite{blog_key}, which introduced a voice activity detection (VAD) model as a preprocessing step. VAD significantly increased the robustness of the overall pipeline, particularly for users with noisy microphones. By filtering out background noise and applying the detection pipeline only when human speech was detected, the system reduced the overall inference volume by approximately 10 percent and improved the quality of inputs. Future improvements in these areas will reduce errors and make AI-based moderation more reliable and effective.

In summary, the Combined Method’s ability to reduce false positives and improve accuracy in identifying non-hate speech made it a valuable enhancement over the GPT Model, aligning well with improving user experience goals and real-world application needs.

\section{LIMITATIONS}
Several limitations in our study might affect the overall effectiveness of the hate speech detection system. 

Firstly, the CNN audio feature model needs more training to enhance its accuracy. The limited number of training data impacts the current model's training. Conducting more training could improve the audio model's prediction effectiveness. 

Additionally, the study encountered challenges due to the insufficiency of training and validating data. The availability of a larger and more diverse dataset could help in developing more effective models and improving overall accuracy. For this, we recognize the need for in-platform (VRChat) data collection, however this imposes challenges regarding data collection and player's privacy that needs to be address.

Lastly, upon manual examination, we discovered background interference, such as music, singing, and laughing, in the audio data reduced the model's detection accuracy. In our study, we did not apply any related noise filtering or preprocessing techniques to mitigate the background interference.

\section{FUTURE WORK}
For conducting our study to improve the hate detection system in the future, several approaches could be considered. 

First, expanding the dataset size would provide a more solid foundation for training the model, which could improve accuracy and generalizability. By incorporating more data samples, especially edge cases, the model would become better at handling diverse scenarios that might otherwise be overlooked.

Second, the current system could be expanded to include multimodal detection by combining both audio and visual inputs in social VR. Incorporating visual and audio data of users' real-time interactions on social VR platforms holds the potential to produce more comprehensive analysis and ultimately enhance detection accuracy. 

Furthermore, adding the capability to identify and categorize various forms of hate speech, such as Racial, Ethnic, Religious, and Sexual Orientation Hate Speech, would significantly augment the model’s detection capabilities. This approach would make the system more effective in recognizing specific types of hateful content, and add reasoning to detection results, thereby increasing its practical usage in social VR.

Additionally, training and assessing the system's performance in specific real-world scenarios such as in real VRChat settings would provide valuable insights into its applicability and robustness.

Lastly, future research should investigate effective human-AI collaboration methods to further reduce false positives and improve the accuracy of hate speech detection by ensuring that automated AI handles routine moderation tasks while human moderators address more complex or nuanced cases.

\section{CONCLUSION}
This study proposes a combined approach for real-time voice-based hate speech detection in social VR that leverages the strengths of both GPT and an audio CNN classifier. The combined method improved precision and reduced false positives, making it more practical for real-world applications. By using GPT as the main method and incorporating the assisting audio classification, the system achieved a more balanced hate speech classification. The findings highlight the potential of multi-modal approaches in enhancing the detection and classification of hate speech, providing a foundation for future improvements and applications in various domains.

\bibliographystyle{abbrv-doi}

\bibliography{safeguard}

\end{document}